\documentclass[english,reprint,showpacs,bibnotes,amsmath,amssymb,aps,floatfix,superscriptaddress]{revtex4-1}
\usepackage{amsmath}
\usepackage{graphicx}

\newcommand{\bvec}{\mathbf}

\begin{document}

\title{Generalized Elliott-Yafet Spin-Relaxation Time for Arbitrary Spin Mixing}

\author{Svenja Vollmar}

\thanks{Also with Graduate School of Excellence \emph{Materials Science in Mainz}, 67663 Kaiserslautern, Germany}

\affiliation{Physics Department and Research Center OPTIMAS, Kaiserslautern University, 67663 Kaiserslautern, Germany}
\author{David J. Hilton}
\affiliation{Department of Physics, The University of Alabama at Birmingham, Birmingham, AL 35294-1170}

\author{Hans Christian Schneider}
\email{hcsch@physik.uni-kl.de}
 
\affiliation{Physics Department and Research Center OPTIMAS, Kaiserslautern University, 67663 Kaiserslautern, Germany}

\date{\today}
\begin{abstract}
We extend our recent result for the spin-relaxation time due to acoustic electron-phonon scattering in degenerate bands with spin mixing [New~J.~Phys.~$\bvec{18}$, 023012 (2015)] to include interactions with optical phonons, and present a numerical evaluation of the spin-relaxation time for intraband hole-phonon scattering in the heavy-hole (HH) bands of bulk GaAs. Comparing our computed spin-relaxation times to the conventional Elliott-Yafet result quantitatively demonstrates that the latter underestimates the spin-relaxation time because it does not correctly describe how electron-phonon interactions change the (vector) spin expectation value of the single-particle states. We show that the conventional Elliott-Yafet spin relaxation time is a special case of our result for weak spin mixing.
\end{abstract}


\maketitle

\section{Introduction} 

The name Elliott-Yafet (EY) mechanism continues to be used in a variety of ways. Originally, it was introduced as the mechanism for spin-relaxation in degenerate bands due to spin-orbit coupling and the electron acoustic-phonon interaction in metals. For this case, Yafet derived an expression for the spin-relaxation time~\cite{Yafet1963}, which we will call the Elliott-Yafet formula and which contains two contributions. One contribution was originally introduced by Elliott, and is due to the combination of any \emph{explicitly spin-independent} scattering process that connects different non-pure spin states~\cite{Elliott1954}. In addition to the Elliott mechanism, there is another contribution to spin relaxation due to the direct modulation of the spin-orbit coupling by the electron-phonon coupling, which was first analyzed by Overhauser~\cite{Overhauser1953}, but is nowadays often called the Yafet contribution~\cite{Xin-Sinova}.

One characteristic of Yafet's formula is that it determines the spin-flip transition probability with the assumption that all states involved are almost pure spin states. That this may be a problem was realized earlier~\cite{Yu2005,Steiauf2010}, but we recently pointed out~\cite{Baral2016} that there is another, potentially more serious problem in that the influence of spin independent scattering processes is not accounted for correctly. As we have argued before~\cite{Baral2016} and show in detail in the present paper, these shortcomings lead, in particular, to an overestimation of the Elliott contribution in Yafet's formula for spin independent scattering processes. In Ref.~\cite{Baral2016} We have derived a new result for the spin-relaxation time in degenerate bands with spin mixing that corrects the shortcomings of Yafet's formula. It is the purpose of the present paper to provide a quantitative comparison between the two expressions. Therefore, we have chosen a model system with a comparatively large spin mixing and efficient electron-phonon coupling. As such an electron-phonon coupling involves optical phonons, we generalize our earlier result for the spin-relaxation time to include different electron-phonon interaction mechanisms. We treat here the spin relaxation of heavy holes in GaAs, in an approximation that makes the hole bands doubly degenerate~\footnote{In our earlier paper~\cite{Baral2016}, we have also called the degenerate bands formed by \emph{non-pure} spin states ``Kramers degenerate,'' which is sometimes done in the literature~\cite{Ferreira-Bastard-92}. Our equation~(2) in~\cite{Baral2016} contained a misprint: The time inversion operator $\hat{K}$ connects states with $\bvec{k}$ and $-\bvec{k}$.}. An advantage of using holes in GaAs is that the electronic states and the material parameters that characterize the electron-phonon interaction are rather well known so that we can achieve an accurate calculation of the spin-relaxation time without the complication of a separate band-structure calculation for the states and the interaction matrix elements. For recent reviews that put hole spin relaxation in the context of semiconductor spintronics, see Ref.~\cite{Wu2010}, and for a general overview of spintronics with an explanation of the Elliott-Yafet mechanism vs.~other spin-relaxation mechanisms, see Refs.~\cite{Zutic2004,Pikus1984}.

We stress that we determine the \emph{spin} dynamics as opposed to calculating transitions rates between opposite pseudo-spin states, which is the quantum number that can be assigned to discriminate between the two degenerate bands~\cite{Ferreira-Bastard-92,Averkiev:2002ti}. We focus on the spin because Yafet's formula has increasingly been used in systems with an equilibrium magnetization~\cite{Steiauf2010,Carva2011,Essert2011}, where the spin dynamics with respect to the fixed magnetization quantization axis is most important. 

Hole spin dynamics after optical excitation in intrinsic bulk GaAs were first measured in Ref.~\onlinecite{Hilton2002}, determining a spin relaxation time for heavy holes of 110\,fs$\pm10\%$ using non-degenerate pump-probe techniques \cite{Hilton2002}. Patz et al. have recently reported hole spin relaxation times in ferromagnetic GaMnAs quantum well of 160--200\,fs.~\cite{Patz2015} We do not attempt a quantitative comparison with these experiments as in Ref.~\cite{Krauss2008,Shen2010}, because we would then also have to account for the dynamics of the optical excitation of electrons and both heavy and light holes~\cite{Dargys:2004ja,Culcer:2006jc}, heavy-hole to light-hole scattering, and the interplay of electron-phonon and electron-electron Coulomb interactions~\cite{Collet:1993ul,Woerner:1995vp}, which makes a straightforward comparison with Yafet's formula difficult, if not impossible.

\section{Model} 

The electron-phonon interaction hamiltonian considered in this paper generalizes the approach of Ref.~\onlinecite{Baral2016} by including more atoms in the unit cell. We highlight the main differences to the derivation in our earlier paper~\cite{Baral2016} and use the same notation. Following Bir and Pikus~\cite{Bir1974}, one writes the change of the potential experienced by the electrons due to phonons in the form 
\begin{equation}
	\delta\hat{v}(\bvec{x})=\sum_{n,r}\bvec{Q}_{n}^{(r)}\cdot\frac{\partial\hat{v}}{\partial\bvec{R}_{n}^{(r)}}(\bvec{x},\{\bvec{R}\}),
\end{equation}
where $\hat v = v(\bvec x)+\xi(\nabla v \times \hat{\bvec{p}}) \cdot \hat{\bvec{s}}$ represents the ionic potential $v(\bvec{x})$ and the spin-orbit interaction with $\xi=\frac{\hbar}{4 m c^2}$.This potential depends on all the ionic coordinates~$\{\bvec{R}\}$. The hat denotes an operator in real space and/or spin space; for instance $\hat{\bvec{p}}=-i\hbar \nabla$, and $\hat{\bvec{s}}$ is the single particle spin operator. Further, $\bvec{R}_{n}$ labels the equilibrium positions of the unit cells in the lattice, $\bvec{R}_{n}^{(r)}$ and $\bvec{Q}_{n}^{(r)}$ the equilibrium position and displacement of the $r$-th atom in the cell at $\bvec{R}_{n}$, respectively. For each phonon branch index~$\lambda$ with phonon polarization vectors $\vec{\varepsilon}\,^{(r)}_{\bvec{q},\lambda}$ we define the operator 
\begin{equation}
	\hat{v}_{\bvec{k}+\bvec{q},\bvec{k}}^{(\lambda)} =- e^{-i(\bvec{k}+\bvec{q})\cdot \bvec{x}}\sum_{n,r} i \ell_{\bvec{q},\lambda}^{(r)} e^{-i\bvec{q}\cdot\bvec{R}_n}
	\vec{\varepsilon}\,^{(r)}_{\bvec{q},\lambda} 
	\cdot \frac{\partial\hat{v}}{\partial\bvec{R}_{n}^{(r)}}e^{i\bvec{k}\cdot\bvec{x}}.
	\label{eq:e-pn-matrix-element}
\end{equation} 
where we have suppressed the dependence of $\hat{v}$ on $(\bvec{x},\{\bvec{R}_{n}\})$. We use the abbreviation $\ell_{\bvec{q},\lambda}^{(r)}=\sqrt{\frac{\hbar}{2 M_r N \omega_{\mathbf{q},\lambda}}}$ where $M_r$ is the mass of the atoms with index $r$ and  where $\omega_{\mathbf{q},\lambda}$ is the phonon dispersion. This allows us to express the electron-phonon matrix element in the form
\begin{equation}
g_{\bvec{k}+\bvec{q}\mu',\bvec{k}\mu}^{(\lambda)}=\langle u_{\bvec{k}+\bvec{q}\mu'}|\hat{v}_{\bvec{k}+\bvec{q},\bvec{k}}^{(\lambda)}u_{\bvec{k}\mu}\rangle,
\label{eq:g-definition}
\end{equation}
where $\mu,\ \mu'$ label the carrier states, and the integrals are extended over the volume~$\Omega$ of the first Brillouin zone. The matrix element enters the electron-phonon interaction hamiltonian 
\begin{equation}
	H_{\mathrm{e-pn}}=\sum_{\bvec{k}\bvec{q}}\sum_{\mu\mu'}\sum_{\lambda}g_{\bvec{k}+\bvec{q}\mu',\bvec{k}\mu}^{(\lambda)}(b_{\bvec{q},\lambda}+b_{-\bvec{q},\lambda}^{\dagger})c_{\bvec{k}+\bvec{q}\mu'}^{\dagger}c_{\bvec{k}\mu}.
	\label{eq:H-e-pn}
\end{equation}

We also define the ($z$ component of the) torque matrix element between Bloch-$u$'s  
\begin{equation}
	t_{\bvec{k}+\bvec{q}\mu',\bvec{k}\mu}^{(\lambda)} = \frac{1}{i\hbar}\big\langle u_{\bvec{k}+\bvec{q}\mu'}|\big[\hat{s}_z,\hat{v}^{(\lambda)}_{\bvec{k}+\bvec{q},\bvec{k}}\big]
	\,u_{\bvec{k}\mu}\big\rangle,
	\label{eq:torque-operator}
\end{equation}
which is a key quantity for the change of spin angular momentum due to incoherent scattering with bosons as described by the interaction hamiltonian~\eqref{eq:H-e-pn}. For an explicitly spin-independent interaction operator $\hat{v}^{(\lambda)}_{\bvec{k}+\bvec{q},\bvec{k}}$ the commutator in Eq.~\eqref{eq:torque-operator} vanishes, and the torque matrix element is zero in any basis. 

\section{Spin-Relaxation Time: New Result vs. Yafet's Formula} 

In Ref.~\onlinecite{Baral2016} we have derived a spin-relaxation time in degenerate bands for a small excited spin polarization, which is the problem originally considered by Yafet \cite{Yafet1963}. Technically, we need to assume a quasi-equilibrium with a prescribed spin polarization (in $z$-direction) as excitation condition and that the Bloch-$u$'s diagonalize the spin operator $\hat{s}_z$ in the degenerate subspace~\cite{Fabian1998,Fabian1999,Cheng2010}, see also Appendix~\ref{sec:Hole-States-GaAs}. Then the spin-relaxation time due to the incoherent scattering with phonons $\tau_{\mathrm{SR}}$ is given by 
\begin{equation}
		\begin{split}\frac{1}{\tau_{\mathrm{SR}}}&=-\frac{2}{\mathcal{N}}\Re\sum_{\bvec{k}\bvec{q}}\sum_{\mu\mu'} \langle s_{z}\rangle_{\bvec{k}+\bvec{q}\mu'}\Delta_{\bvec{k}+\bvec{q}} 		\\
			 &\times \sum_{\lambda}t_{\bvec{k}+\bvec{q}\mu',\bvec{k}\mu}^{\left(\lambda\right)}g_{\bvec{k}\mu,\bvec{k}+\bvec{q}\mu'}^{(\lambda)}
	 \Big(\frac{1+N_{\bvec{q},\lambda}-n_{\bvec{k}}}{\epsilon_{\bvec{k}+\bvec{q}}-\epsilon_{\bvec{k}}-\hbar\omega_{\bvec{q},\lambda}+i\hbar\gamma}\\
			&\phantom{\times \sum_{\lambda}t_{\bvec{k}+\bvec{q}\mu',\bvec{k}\mu}^{\left(\lambda\right)}g_{\bvec{k}\mu,\bvec{k}+\bvec{q}\mu'}^{(\lambda)}} 
			+\frac{N_{\bvec{q},\lambda}+n_{\bvec{k}}}{\epsilon_{\bvec{k}+\bvec{q}}-\epsilon_{\bvec{k}}
				+\hbar\omega_{\bvec{q},\lambda}+i\hbar\gamma}\Big).
		\end{split}
	\label{eq:spin-relaxation-time}
\end{equation}
Here and in the following we use the notation 
\begin{equation}
\langle s_z\rangle_{\bvec{k}\mu} =\langle u_{\bvec{k}\mu}|\hat{s}_{z}u_{\bvec{k}\mu}\rangle
\end{equation}
 for the single-particle spin expectation value. The occupation number of the phonon bath is described by the Bose function~$N_{\bvec{q},\lambda} = b\left(\hbar\omega_{\bvec{q},\lambda}\right)$ while the carrier distribution is a Fermi-Dirac distribution $n_{\bvec{k}}=f(\epsilon_{\bvec{k}}-\mu)$, with the chemical potential $\mu$. For small temperatures, the function $\Delta_{\bvec{k}}=-\frac{\partial f}{\partial\epsilon}\big|_{\epsilon_{\bvec{k}}-\mu}$ approaches a $\delta$-function in energy peaked at the chemical potential $\mu$, and $\hbar\gamma$ is an infinitesimal broadening. By $\epsilon_{\bvec{k}}$ we denote the dispersion of the degenerate pair of bands while the phonon dispersion is given by $\hbar\omega_{\bvec{q},\lambda}$. The normalization factor is $\mathcal{N} =\sum_{\bvec{k}}\sum_{\mu}|\langle s_{z}\rangle_{\bvec{k}\mu}|^{2}\Delta_{\bvec{k}}$, which is related to a Fermi surface average of the squared $\hat{s}_z$ expectation value
\begin{equation} 
	\bar{s}_z=\sqrt{\frac{\sum_{\mu}\sum_{\bvec{k}}|\langle s_{z}\rangle_{\bvec{k}\mu}|^{2}\Delta_{\bvec{k}}}{\sum_{\mu}\sum_{\bvec{k}}\Delta_{\bvec{k}}}}
	\label{eq:s-squared-average}
\end{equation}
via $\mathcal{N}=(\sum_{\mu,\bvec{k}}\bar{s}_z\Delta_{\bvec{k}})^2$. 

Next, we show the approximations one has to employ in Eq.~\eqref{eq:spin-relaxation-time} to recover Yafet's formula~\cite{Yafet1963}. This provides an important check on our result~\eqref{eq:spin-relaxation-time}. Starting from Eq.~\eqref{eq:spin-relaxation-time} by explicitly using the expression for the torque matrix element, see Eq.~\eqref{eq:torque-operator}, we find 
\begin{widetext}
\begin{equation}
		\begin{split}\frac{1}{\tau_{\mathrm{SR}}}
			=\frac{2}{\mathcal{N}\hbar}\Re\sum_{\bvec{k}\bvec{q}}\sum_{\mu\mu',\lambda} & i\big[\langle s_{z}\rangle_{\bvec{k}+\bvec{q}\mu'}-\langle s_{z}\rangle_{\bvec{k}\mu}\big]\big|\langle u_{\bvec{k}\mu}|\hat{v}_{\bvec{k},\bvec{k}+\bvec{q}}^{(\lambda)}u_{\bvec{k}+\bvec{q}\mu'}\rangle\big|^{2}\langle s_{z}\rangle_{\bvec{k}+\bvec{q}\mu'}\Delta_{\bvec{k}+\bvec{q}}\\
			\times & \left(\frac{1+N_{\bvec{q},\lambda}-f_{\bvec{k}}}{\epsilon_{\bvec{k}+\bvec{q}}-\epsilon_{\bvec{k}}-\hbar\omega_{\bvec{q},\lambda}+i\hbar\gamma}+\frac{N_{\bvec{q},\lambda}+f_{\bvec{k}}}{\epsilon_{\bvec{k}+\bvec{q}}-\epsilon_{\bvec{k}}+\hbar\omega_{\bvec{q},\lambda}+i\hbar\gamma}\right),
		\end{split}
	\end{equation}
\end{widetext}
where we used the completeness relation $\sum_{\nu}|u_{\bvec{k}\nu}\rangle\langle  u_{\bvec{k}\nu}|=\mathbb{1}$ and the following property of the eigenbasis $\langle u_{\bvec{k}\mu}|\hat{s}_{z}|u_{\bvec{k}\nu}\rangle=\langle u_{\bvec{k}\mu}|\hat{s}_{z}|u_{\bvec{k}\mu}\rangle\delta_{\mu\nu}$. The latter relation is guaranteed by construction of the $|u_{\bvec{k}\mu}\rangle$ states, see the discussion in Appendix~\ref{sec:Hole-States-GaAs} following Eq.~\eqref{eq:u-spin-mixing}.

Now, we have to assume that only \emph{interband} scattering processes can change the spin, i.e. $\mu'=\bar{\mu}$. This is Yafet's assumption that applies rigorously to purely spin up and spin down bands, but is an approximation for the case of $\bvec{k}$-dependent spin mixing. In the spirit of Yafet's assumption, we also approximate the matrix elements of the spin operator by the values~$\pm\frac{\hbar}{2}$, i.e., we make the replacements
\begin{equation}
	\left(\langle s_{z}\rangle _{\bvec{k}+\bvec{q}\bar{\mu}}-\langle s_{z}\rangle_{\bvec{k}\mu}\right)\langle s_{z}\rangle_{\bvec{k}+\bvec{q}\bar{\mu}}
		=\frac{\hbar^{2}}{4}-\left(-\frac{\hbar^{2}}{4}\right)=\frac{\hbar^{2}}{2}
\end{equation}
and $\mathcal{N}=\sum_{\bvec{k'}}\sum_{\mu}\big|\langle s_{z}\rangle _{\bvec{k'}\mu}\rangle\big|^{2}\Delta_{\bvec{k'}}=(\hbar^{2}/2)\sum_{\bvec{k'}}\Delta_{\bvec{k'}}$.
With $\Re(ia)=-\Im(a)$ this leads to
\begin{widetext}
	\begin{equation}
		\frac{1}{\tau_{\mathrm{SR}}}=-\frac{2}{\hbar\sum_{\bvec{k'}}\Delta_{\bvec{k'}}}\Im\sum_{\bvec{k}\bvec{q}}\sum_{\mu,\lambda}  \big|g^{(\lambda)}_{\bvec{k}\mu,\bvec{k}+\bvec{q}\bar{\mu}}\big|^{2}\Delta_{\bvec{k}+\bvec{q}}\left(\frac{1+N_{\bvec{q},\lambda}-f_{\bvec{k}}}{\epsilon_{\bvec{k}+\bvec{q}}-\epsilon_{\bvec{k}}-\hbar\omega_{\bvec{q},\lambda}+i\hbar\gamma}+\frac{N_{\bvec{q},\lambda}+f_{\bvec{k}}}{\epsilon_{\bvec{k}+\bvec{q}}-\epsilon_{\bvec{k}}+\hbar\omega_{\bvec{q},\lambda}+i\hbar\gamma}\right).
	\label{eq:Yafet-our-form}
	\end{equation}
\end{widetext}
where we have also inserted the definition of the matrix element~\eqref{eq:g-definition}. 

Equation~\eqref{eq:Yafet-our-form} is essentially Yafet's Eq.~(18.6) in Ref.~\cite{Yafet1963} in our notation. For completeness, we note the steps to get the exact analogy to his result. He uses the Dirac identity 
\begin{equation}
		\Im\frac{1}{\Delta E+i\hbar\gamma}\overset{\gamma\rightarrow0}{\rightarrow}-\pi\delta\left(\Delta E\right)
\label{eq:Dirac}
\end{equation}
and replaces 
\begin{equation}
\sum_{\mu}\big|g_{\bvec{k}\mu,\bvec{k}+\bvec{q}\bar{\mu}}\rangle\big|^{2}
		 =2\frac{\hbar}{2\rho V^{2}\omega_{\bvec{q},\lambda}}\left|M_{\bvec{k}+\bvec{q}\Uparrow,\bvec{k}\Downarrow}^{\left(\lambda\right)}\right|^{2}.
\end{equation}
Note that relation~\eqref{eq:Dirac} removes the misleading minus sign in front of the relaxation time. This concludes the proof that Yafet's result~\eqref{eq:Yafet-our-form} is a special case of our result~\eqref{eq:spin-relaxation-time} for weak spin mixing.

\section{States and Hole-Phonon Interaction} 

We now turn to the numerical evaluation of the spin-relaxation time~\eqref{eq:spin-relaxation-time}. To this end, we need the states $u_{\bvec{k}\mu}$, the band dispersions, electron-phonon interaction and torque matrix elements as well as the expectation values of the spin operator $\hat{s}$. For our chosen model system of heavy holes in GaAs, these band dispersions and matrix elements are readily available. Since we want to treat the relaxation of a small spin polarization created in an equilibrium population of electrons, we analyze the case of a relaxed population of heavy holes in this band structure, as they would be created in $p$-doped GaAs, but we neglect the influence of the dopant ions. This has the further advantage that the Fermi surface of the relaxed heavy holes is comparatively simple. With these assumptions, we have a test bed for our spin-relaxation time that is well defined, and can be checked independently with a limited numerical effort that does not require \emph{ab initio} calculations. Thus, our model for the band structure is the $4\times 4$ Luttinger hamiltonian with standard parameters for GaAs taken from Ref.~\onlinecite{Chuang1995}. This hamiltonian does not include $k^3$ terms, which give rise to spin splitting due to the bulk inversion asymmetry in GaAs. In this approximation, the hole bands become degenerate at every $\mathbf{k}$-point and realize the conditions for which Yafet's result was derived.

We choose the degenerate eigenstates as the two states that diagonalize the spin operator $\hat{s}_{z}$ in the two-dimensional subspace of HH bands at each $\bvec{k}$~\cite{Fabian1998,Fabian2007,Baral2016}. The explicit form of the Luttinger hamiltonian and the states are given in Appendix~\ref{sec:Hole-States-GaAs}. To quantify the spin mixing in these states, see Eq.~\eqref{eq:superintelligent-basis}, we note that the single-particle spin averaged over the Fermi surface, as defined in~\eqref{eq:s-squared-average}, is $\bar{s}_z=\pm 0.58\times\frac{\hbar}{2}$. Yafet's derivation assumes that this value is $\bar{s}=\pm\frac{\hbar}{2}$. It is a peculiarity of the hole states considered here that this value is independent of the chemical potential and thus the same for all hole densities.

For the interaction of electrons with phonons, one can distinguish, in general, contributions that are spin-independent and explicitly spin-dependent, as well as those due to short-range and long-range interactions. We discuss the physics here and present the details in Appendix~\ref{sec:hole-phonon-interaction}. For acoustic phonons, there are no long-range contributions, and we determine the matrix element in the long-wavelength limit following Bir and Pikus~\cite{Bir1974}. Importantly, this matrix element, see Eq.~\eqref{eq:acoustic-matrixelement-D}, yields a non-vanishing torque matrix element because it includes both spin-independent and explicitly spin-dependent contributions. 

For longitudinal optical (LO) phonons, there is a short-range and a long-range contribution, so that the interaction operator takes the form $\hat{v}^{(\text{LO})}=\hat{v}^{(\text{LO,lr})}+\hat{v}^{(\text{LO,sr})}$. The long-range part of the interaction is usually called Fr\"{o}hlich, or polar, coupling, and is due to long-range electrostatic fields set up by the vibrating ions. This is the most effective coupling for momentum relaxation, as its matrix element behaves like $g^{(\text{LO,lr})} \propto q^{-1}$, see, e.g., Ref.~\onlinecite{Mahan2000}. However, it is electrostatic in nature, i.e., $\hat{v}^{(\text{LO,lr})}$ is explicitly spin independent, so that its torque $[\hat{s}_z,\hat{v}^{(\text{LO,lr})}_{\bvec{k}+\bvec{q},\bvec{k}}] $, and all of the matrix elements~\eqref{eq:torque-operator}, regardless of the basis, vanish~\cite{Scholz1995}.

The short-range interaction, which is sometimes called the nonpolar optical phonon coupling or deformation potential interaction exists for all three optical phonon branches; it has a matrix element $g^{(\lambda,\text{sr})}$ (where $\lambda$ labels the optical phonon branch), see Eq.~\eqref{eq:deformation-potential-optical-phonons}, that is independent of $q$ for $q \to 0$~\cite{Scholz1995, Bir1974}. In contrast to the long-range polar interaction, the short-range nonpolar interaction is explicitly spin dependent, and it therefore represents the most effective spin-dependent electron-phonon interaction in GaAs. This was realized in Ref.~\cite{Shen2010}, but not included in earlier attempts at calculating hole-spin relaxation after optical excitation~\cite{Yu2005,Krauss2008}. 

\begin{figure}[t]
	\begin{centering}
		\includegraphics[width=1\linewidth]{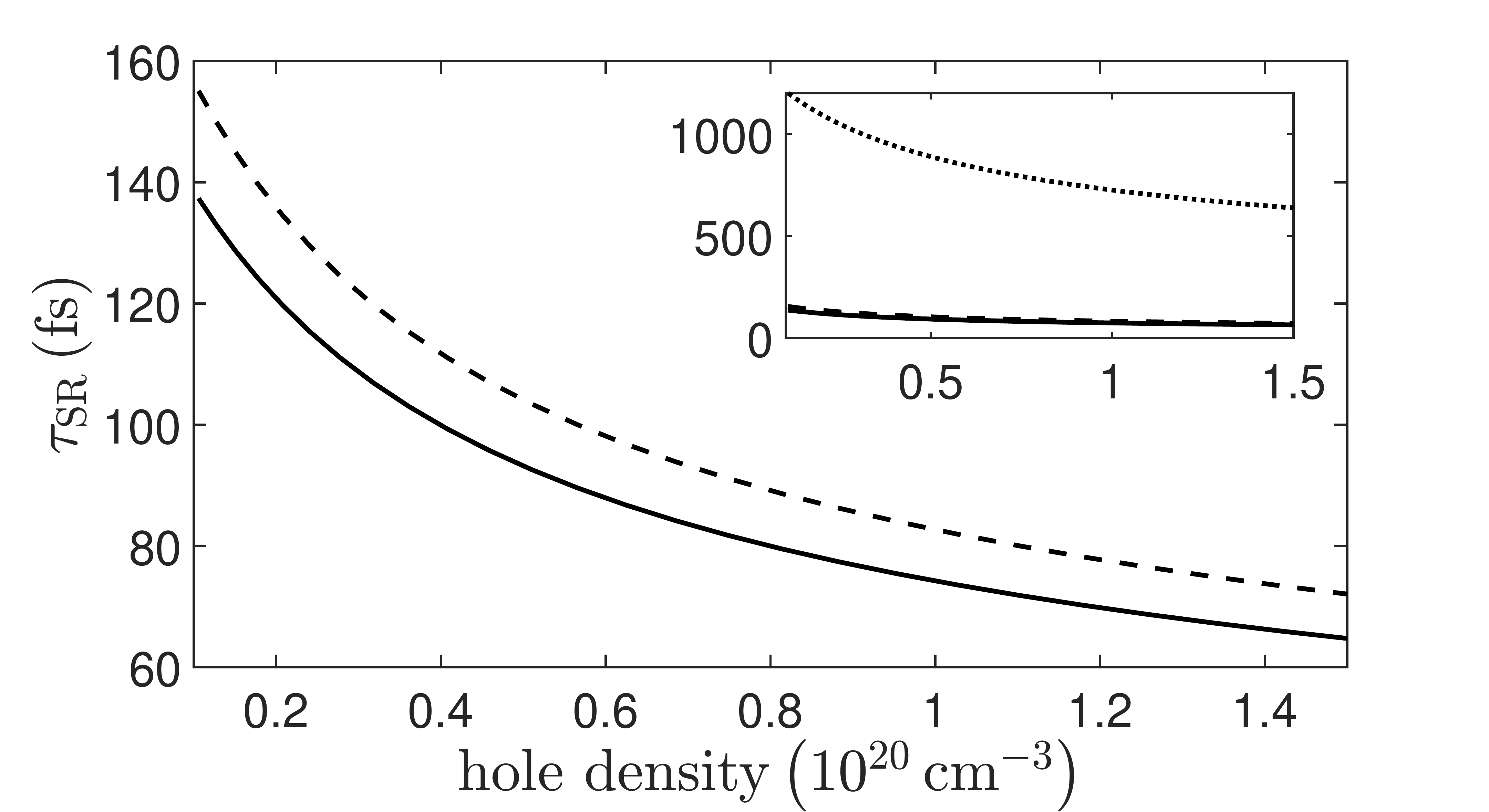} 
		\par\end{centering}
	\caption{Density dependent HH intraband spin relaxation time calculated using Eq.~\eqref{eq:spin-relaxation-time} for $T_{\mathrm{eq}}=300$\,K. The solid line includes all coupling mechanisms, the dashed line only the contribution of the nonpolar coupling to optical phonons (dashed), and dotted line (inset) the deformation potential coupling to acoustic phonons.}
	\label{fig:our_density_dependence} 
\end{figure}

\section{Numerical Results} 

We now discuss results for the spin-relaxation time computed with Eq.~\eqref{eq:spin-relaxation-time} using the states and matrix elements for heavy holes of GaAs. We treat the case of a relaxed density of holes, and do not model the creation of these holes by optical excitation, or effects of doping. The hole distribution is thus described by a Fermi-Dirac distribution with a chemical potential~$\mu$ determined by the density and an electronic temperature~$T_{\mathrm{eq}}=300$\,K that equals the temperature of the phonon bath. The numerical broadening $\hbar\gamma$ in Eq.~\eqref{eq:spin-relaxation-time} is chosen to be 0.6\,meV and we have checked that the results are converged. We consistently work in the long-wavelength limit and follow Scholz~\cite{Scholz1995} in choosing Cartesian polarization vectors for the long-wavelength optical phonons; we therefore have to approximate the different LO and TO phonon energies by an average value.

In our Eq.~\eqref{eq:spin-relaxation-time}, and also Yafet's result~\eqref{eq:Yafet-our-form}, the contributions from different phonon modes are added. For LO phonons, the product $(g^{(\text{lr})}+g^{(\text{sr})})t^{(\text{sr})}$ occurs in Eq.~\eqref{eq:spin-relaxation-time}, because there is no long-range contribution to the torque. Due to the particular form of the matrix elements for holes in the long-wavelength limit, the cross terms of the form $t^{(\text{lr})}\,g^{(\text{sr})}$ vanish after integration over $\bvec{q}$, so that we  effectively have three additive contributions to the spin-relaxation time from Eq.~\eqref{eq:spin-relaxation-time}, and to Yafet's original relation~\eqref{eq:Yafet-our-form}: the deformation potential coupling to acoustic phonons, the non-polar coupling to optical phonons and the polar coupling to LO phonons. We compare their respective contributions in the following.

Figure~\ref{fig:our_density_dependence} shows the total spin-relaxation time $\tau_{\text{SR}}$ (solid line) calculated with our Eq.~\eqref{eq:spin-relaxation-time} as well as the contributions from the different couplings. The dominant one is the nonpolar coupling to optical phonons (dashed line), with the the acoustic phonons having only a very small effect over the whole range of hole densities considered here. As expected, $\tau_{\text{SR}}$ decreases with hole density.

\begin{figure}[t]
	\includegraphics[width=1\linewidth]{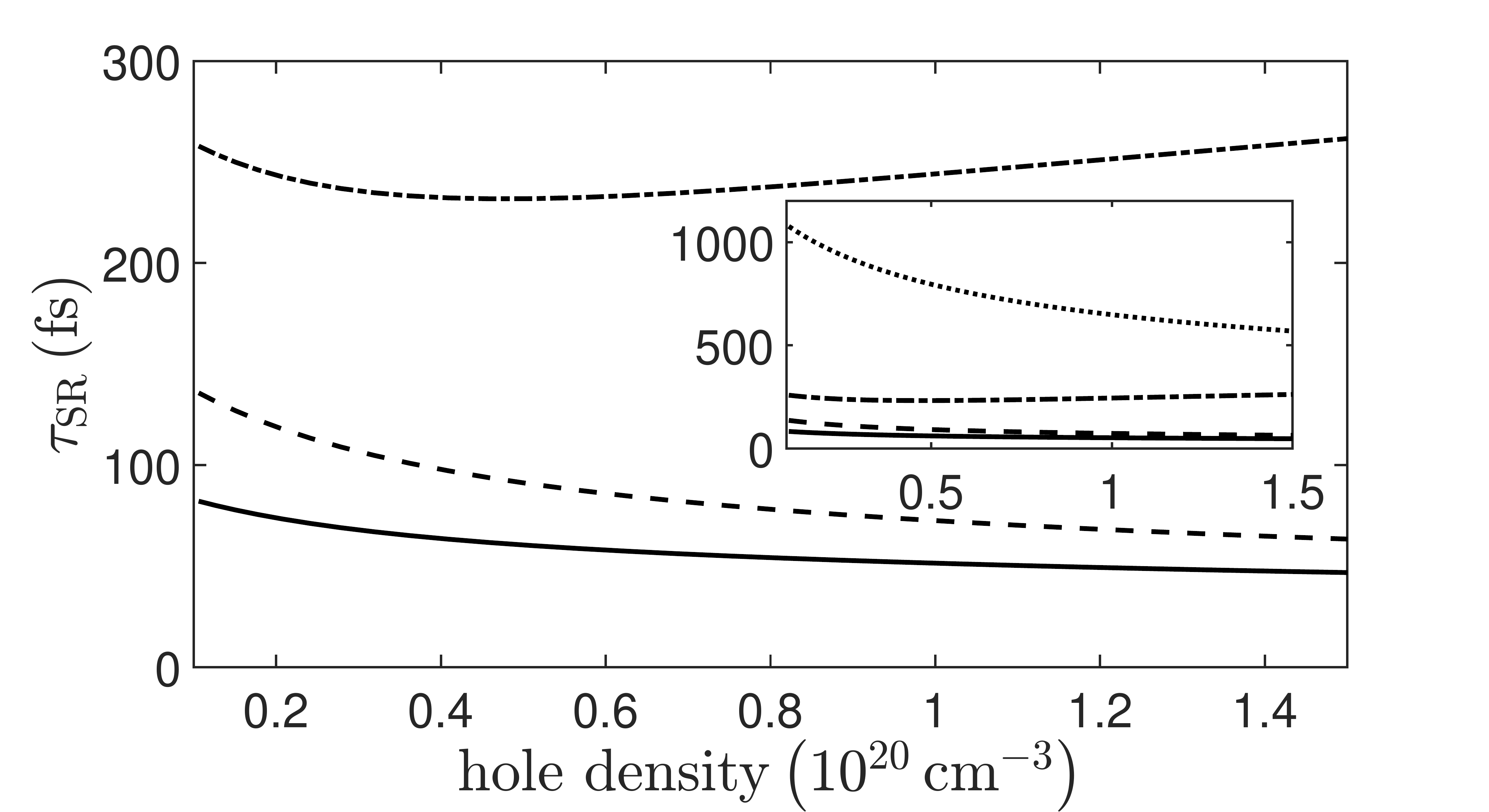} 
	\caption{Density dependent HH intraband spinrelaxation time at $T_{\mathrm{eq}}=300$\,K calculated using Yafet's formula. Shown are the different contributions of non nonpolar phonons (dashed line), polar (dashed-dotted) and acoustic (dotted, Inset) phonons. The complete result ist given by the solid line. 
		\label{fig:Yafet_density_dependence} }
\end{figure}

In Fig.~\ref{fig:Yafet_density_dependence} we use Yafet's original relation~\eqref{eq:Yafet-our-form}, to compute the spin-relaxation time and the different contributions to it for the same parameters as in Fig.~\ref{fig:our_density_dependence}. Note first that there is now also a contribution from the polar coupling to LO phonons. Although the nonpolar coupling to optical phonons is the dominant contribution, the polar coupling is on the same order of magnitude. For larger hole densities the influence of the polar coupling decreases, and the acoustic phonons do not contribute much over the whole density range studied here. When comparing the ratio between the polar and nonpolar contributions, one needs to keep in mind that the contribution from the nonpolar coupling, and thus also the total spin-relaxation times, depend on the magnitude of the optical deformation potential constant $d_{0}$. Different values for this parameter have been reported, ranging from 27.4\,eV \cite{Scholz1995}, over 36.4\,eV~\cite{Blacha1984} to experimentally deduced values of 48\,eV~\cite{Poetz1981,Grimsditch1979}. Since $d_{0}=48$\,eV was also used in other theoretical calculations~\cite{Shen2010,Langot1996} we choose this value, but if the true value is closer to 28\,eV, then the \emph{polar} optical phonon coupling becomes the dominant contribution in Yafet's formula~\eqref{eq:Yafet-our-form}, which is a qualitatively wrong result. 

\begin{figure}[t]
	\includegraphics[width=1\linewidth]{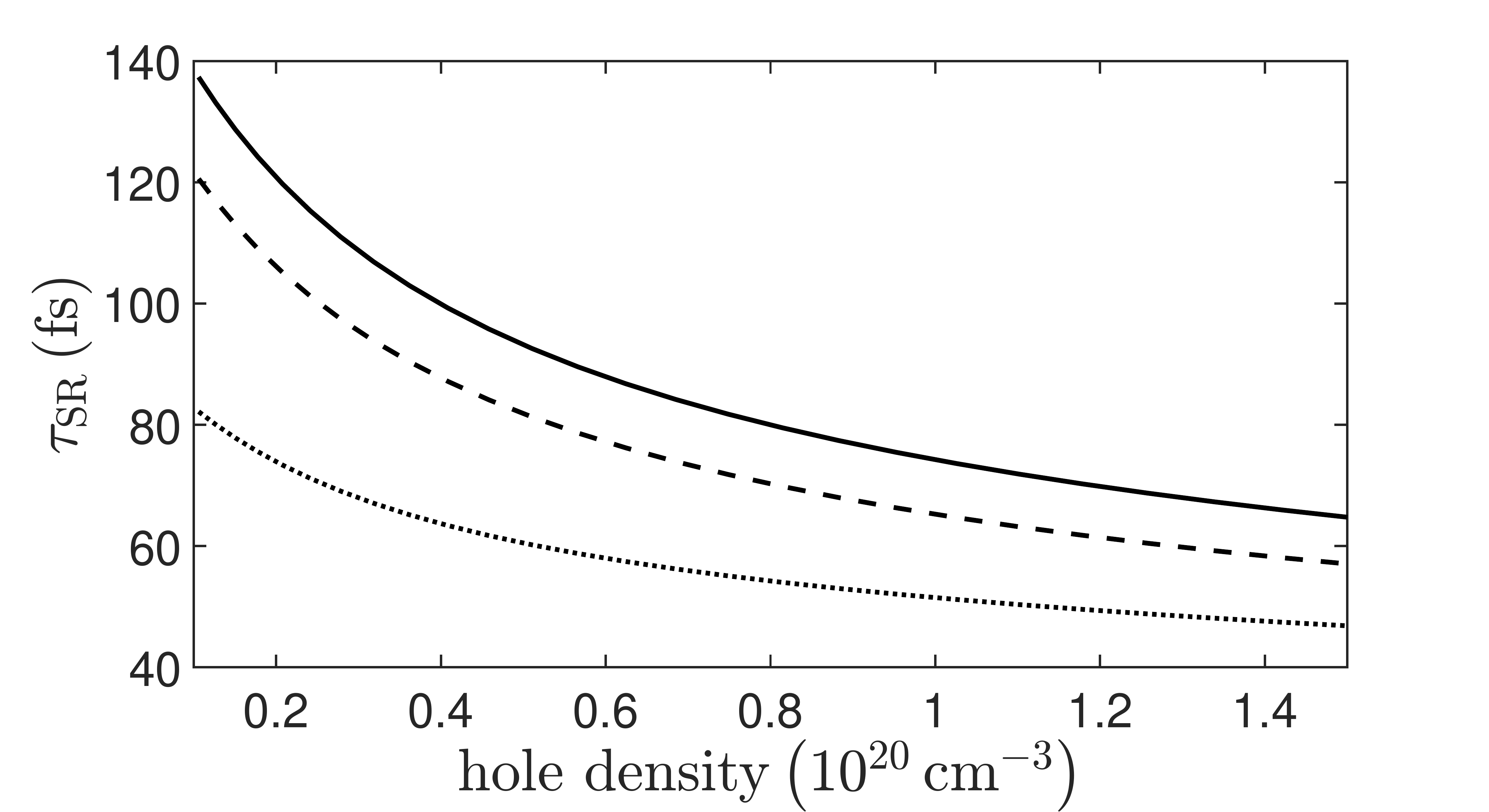} 
	\caption{Comparison of spin-relaxation times computed with Eq.~\eqref{eq:spin-relaxation-time} (solid line) and Yafet's formula for two cases: with (dotted) and without (dashed) the spin-independent polar coupling to optical phonons. In Eq.~\eqref{eq:spin-relaxation-time} the polar coupling does not contribute because it has a vanishing torque matrix element.~\label{fig:density-dependence-comparison}}
\end{figure}

Figure~\ref{fig:density-dependence-comparison} compares the most important features of the two different calculations presented in Figs.~\ref{fig:our_density_dependence} and~\ref{fig:Yafet_density_dependence}. The solid line represents the complete spin-relaxation time calculated using Eq.~\eqref{eq:spin-relaxation-time}. For Yafet's formula we show two results: one with all electron-phonon coupling mechanisms included (dotted line) and one without the polar coupling (dashed line). Including all coupling mechanisms (dotted line) leads to a spin-relaxation time that is off by about 35\% or more for the density range considered here. Note that the absolute value of the deviation depends on $d_{0}$. This result shows how Yafet's formula deviates from Eq.~\eqref{eq:spin-relaxation-time}, which contains a correct description of the single-particle spin expectation values at each $k$ point, and the change of the spin vector due to scattering transitions. 

An interesting comparison results if we investigate numerically the influence of the polar (or Fr\"{o}hlich) coupling, as shown in Fig.~\ref{fig:density-dependence-comparison}. The polar coupling to LO phonons does not contribute at all in our result~\eqref{eq:spin-relaxation-time} due to its vanishing torque matrix element. However, its interaction matrix element $g^{\text{lr}}$ connects states at $\bvec{k}$ and $\bvec{k}+\bvec{q}$ with different spin expectation values, so that switching it off in Yafet's formula changes the spin-relaxation time; numerically we find for the case without the spin-independent Fr\"{o}hlich interaction a reduced deviation of about~14\%, independent of the hole density. Thus  Yafet's formula~\eqref{eq:Yafet-our-form} gives a better result by neglecting the spin-independent polar LO-phonon interaction, even though this is the most efficient momentum scattering mechanism. Stated differently, Fig.~\ref{fig:density-dependence-comparison} shows that \emph{Yafet's formula~\eqref{eq:Yafet-our-form} may massively overestimate the contribution of an efficient spin-independent scattering process, the influence of which on the spin dynamics is usually called the Elliott mechanism}. The remaining difference on the order of 10\% between the dashed line and our full result is due to interaction mechanisms that contain both spin-independent and spin-dependent contributions. Given that Yafet's assumption of the spin expectation value $\bar{s}_z=\pm \frac{\hbar}{2}$ is violated badly in the present system and the heavy-hole states do not at all resemble pure spin states, this relatively small remaining deviation may at first be surprising. However, one needs to keep in mind that our result (as opposed to Yafet's) includes a correct determination of the ensemble spin, even for pronounced spin mixing. It turns out that the spin changing transitions are interband transitions, both in our treatment and Yafet's. The difference is that in Yafet's treatment those transitions connect almost pure spin states and therefore flip an $\hbar/2$ spin. In our case they change the spin expectation value by a smaller amount, but this smaller change occurs with respect to an ensemble spin determined for non-pure spin states. On average, when spin mixing is included correctly, a single transition in our treatment needs to flip ``less spin'' than in Yafet's calculation. 

\section{Conclusion} 

We extended our analysis of spin relaxation due to electron-phonon interactions~\cite{Baral2016} to include coupling to optical phonons, which may have long-range electrostatic interactions. Our approach accounts for the vector spin expectation values at each $k$ point and correctly describes how different electron-phonon interactions change the spin in scattering transitions. We showed that Yafet's result is a special case of the spin-relaxation time derived by us. We applied our expression for the spin-relaxation time in degenerate bands with spin mixing to the test case of phonon scattering in heavy-hole bands in GaAs. By computing the spin-relaxation time for different hole densities, we quantitatively showed that Yafet's result leads to a 35\% shorter spin-relaxation time at low hole densities than our spin-relaxation time. The biggest difference is that in Yafet's result the long-range LO phonon coupling contributes to spin relaxation, whereas in our calculation this spin-independent interaction does not contribute at all. Yafet's formula therefore tends to overestimate the contribution to spin relaxation of efficient spin-independent scattering mechanisms.

\begin{acknowledgments}
	We acknowledge support from the DFG through the SFB/TRR 173 ``Spin+X'' (Project A8). Svenja Vollmar was supported by the Excellence Initiative (DFG/GSC 266).
\end{acknowledgments}

\begin{appendix}

\section{Hole States in GaAs\label{sec:Hole-States-GaAs}}

To calculate the intraband HH spin-relaxation time, we use a $4\times4$
Luttinger hamiltonian which describes the heavy-hole and light-hole
states close to the fundamental band gap in GaAs. For our purposes,
it is enough to include HH and LH without coupling to the split-off
and electronic bands. The representation of the hamiltonian in the
$k=0$ eigenstates is given by \cite{Chuang1995}
\begin{equation}
	\hat{h}_{\mathrm{hole}}=\begin{pmatrix}P+Q & -S & R & 0\\
		-S^{*} & P-Q & 0 & R\\
		R^{*} & 0 & P-Q & S\\
		0 & R^{*} & S^{*} & P+Q
	\end{pmatrix},
\end{equation}
with the abbreviations 
\begin{equation}
	P=  \frac{\hbar^{2}}{2m_{0}}\gamma_{1}\left(k_{z}^{2}+k_{\perp}^{2}\right), \quad
		Q=  \frac{\hbar^{2}}{2m_{0}}\gamma_{2}\left(-2k_{z}^{2}+k_{\perp}^{2}\right)
\end{equation}
\begin{equation}
		S= \frac{\hbar^{2}}{2m_{0}}2\sqrt{3}\gamma_{3}k_{z}k_{-},
\end{equation}
and
\begin{equation}		
	R=  -\frac{\hbar^{2}}{2m_{0}}\sqrt{3}\gamma_{2}\left(k_{x}^{2}-k_{y}^{2}\right)+\frac{\hbar^{2}}{2m_{0}}2i\sqrt{3}\gamma_{3}k_{x}k_{y}.
\end{equation}
We write $k_{\pm}=k_{x}\pm ik_{y}$ and $k_{\perp}^{2}=k_{x}^{2}+k_{y}^{2}$ and use standard parameters $\gamma_1=6.85$, $\gamma_2=2.1$ and $\gamma_3 = 2.9$.

The twofold degenerate eigenenergies are 
\begin{equation}
	\begin{split}\epsilon_{\mathrm{HH}} & =P-\sqrt{Q^{2}+\left|R\right|^{2}+\left|S\right|^{2}},\\
		\epsilon_{\mathrm{LH}} & =P+\sqrt{Q^{2}+\left|R\right|^{2}+\left|S\right|^{2}},\label{eq:eigenenergies}
	\end{split}
\end{equation}
and the eigenstates can be expressed in the form \cite{Chuang1995}
\begin{widetext}
	\begin{equation}
		\begin{split}|\mathrm{HH},1\rangle & =\begin{pmatrix}S\\
				Q+\sqrt{Q^{2}+\left|R\right|^{2}+\left|S\right|^{2}}\\
				0\\
				-R^{*}
			\end{pmatrix},\ |\mathrm{HH},2\rangle=\begin{pmatrix}-R\\
				0\\
				Q+\sqrt{Q^{2}+\left|R\right|^{2}+\left|S\right|^{2}}\\
				-S^{*}
			\end{pmatrix},\\
			|\mathrm{LH},1\rangle & =\begin{pmatrix}Q+\sqrt{Q^{2}+\left|R\right|^{2}+\left|S\right|^{2}}\\
				-S^{*}\\
				R^{*}\\
				0
			\end{pmatrix},\ |\mathrm{LH},2\rangle=\begin{pmatrix}0\\
				-R\\
				-S\\
				Q+\sqrt{Q^{2}+\left|R\right|^{2}+\left|S\right|^{2}}
			\end{pmatrix}.\label{eq:Kramers-konjugierteBasis1}
		\end{split}
	\end{equation}
\end{widetext}

We will illustrate here a detail of the calculation for the Kramers-degenerate eigenstates of the Luttinger hamiltonian, which is rarely mentioned in the semiconductor-literature, but has some importance for spin relaxation. In order to quantifiy the ``spin mixing'', one would like to write any state involved in the dynamics in the form
\begin{equation}
	|u_{\bvec{k}\mu}\rangle=a_{\bvec{k}\mu}|\uparrow\rangle+b_{\bvec{k}\mu}|\downarrow\rangle.
\label{eq:u-spin-mixing}
\end{equation}
This decomposition is not unique because two eigenstates are degenerate.
Consequently, there are different explicit expressions for the heavy-hole~$|\text{HH},1\rangle$,
$|\text{HH,2\ensuremath{\rangle}}$ and light-hole eigenstates in
the literature. To fix the the eigenstates and the spin mixing parameters $a_{\bvec{k}\mu}$ and $b_{\bvec{k}\mu}$, we choose the Kramers conjugate eigenbasis $\{|\widetilde{\mathrm{HH}}1\rangle, |\widetilde{\mathrm{HH}}2\rangle , |\widetilde{\mathrm{LH}}1\rangle , |\widetilde{\mathrm{LH}}2\rangle  \}$ that diagonalizes the spin operator in the subspace of the HH and LH eigenstates. The representation of the spin operator in the $k=0$ eigenstates is given by 
\begin{equation}
	\hat{s}_{z}=\frac{\hbar}{2}\begin{pmatrix}1 & 0 & 0 & 0\\
		0 & \frac{1}{3} & 0 & 0\\
		0 & 0 & -\frac{1}{3} & 0\\
		0 & 0 & 0 & -1
	\end{pmatrix}.
\end{equation}
Only if 
\begin{equation}
	\langle\widetilde{\mathrm{HH}},1|\hat{s}_{z}|\widetilde{\mathrm{HH}},2\rangle=\langle\widetilde{\mathrm{HH}},1|\hat{s}_{z}|\widetilde{\mathrm{HH}},1\rangle=0
\end{equation}
is fulfilled the spin mixing parameter is meaningful \cite{Fabian2007,Fabian1998,Fabian1999,Cheng2010}.
In this way, the Kramers conjugate eigenstates for the HH bands (for
LH replace subscripts HH by LH) are given by superpositions of the
states usually used in semicondcutor physics; in detail
\begin{equation}
	\begin{split}|\widetilde{\mathrm{HH},1}\rangle & =\frac{1}{\mathcal{N}_{\mathrm{HH},1}}\left[|\mathrm{HH},1\rangle+b_{\mathrm{HH}}|\mathrm{HH},2\rangle\right],\\
		|\widetilde{\mathrm{HH},2}\rangle & =\frac{1}{\mathcal{N}_{\mathrm{HH},2}}\rangle\left[|\mathrm{HH},2\rangle-b_{\mathrm{HH}}^{*}|\mathrm{HH},1\rangle\right].
	\end{split}
	\label{eq:superintelligent-basis}
\end{equation} 
Here, the abbreviations 
\begin{equation}
	\begin{split}b_{\mathrm{HH}} & =-\frac{SR^{*}}{2\left|SR\right|^{2}}\left(D_{\mathrm{HH}}-\sqrt{D_{\mathrm{HH}}^{2}+4\left|SR\right|^{2}}\right),\\
		b_{\mathrm{LH}} & =-\frac{3S^{*}R^{*}}{2\left|SR\right|^{2}}\left(D_{\mathrm{LH}}-\sqrt{D_{\mathrm{LH}}^{2}+\frac{4}{9}\left|SR\right|^{2}}\right),
	\end{split}
\end{equation}
with 
\begin{equation}
	\begin{split}D_{\mathrm{HH}} & =\frac{1}{3}\left(Q+\sqrt{Q^{2}+\left|R\right|^{2}+\left|S\right|^{2}}\right)^{2}-\left|R\right|^{2}+\left|S\right|^{2},\\
		D_{\mathrm{LH}} & =\left(Q+\sqrt{Q^{2}+\left|R\right|^{2}+\left|S\right|^{2}}\right)^{2}+\frac{1}{3}\left|R\right|^{2}-\frac{1}{3}\left|S\right|^{2}
	\end{split}
\end{equation}
are used. $\mathcal{N}_{\mathrm{HH}/\mathrm{LH},1/2}$ is a
normalization factor for the mixed eigenstates.

\section{The hole-phonon interaction\label{sec:hole-phonon-interaction}}

The derivation of the spin-relaxation time as given in Eq.~(5)
is based on the general electron-phonon interaction hamiltonian~(3)
with the interaction matrix element $g_{\bvec{k}+\bvec{q}\mu',\bvec{k}\mu}^{(\lambda)}$,
which includes contributions from all the relevant electron-phonon
coupling mechanism. In the case of the polar semiconductor GaAs, we
need to consider a mechanism that is due to long-range dipolar electrostatic
field, which gives rise to the Fr\"{o}hlich coupling to LO phonons, and
the short-range deformation potential interaction \cite{Scholz1995}.

The Fr\"{o}hlich coupling is given by 
\begin{widetext}
	\begin{equation}
		\left|v_{\bvec{k}+\bvec{q}\mu,\bvec{k}\mu'}^{(\lambda)}\right|^{2}=\left|\langle u_{\bvec{k}+\bvec{q}\mu}|\hat{v}_{\bvec{k}+\bvec{q},\bvec{k}}^{(\lambda)}|u_{\bvec{k}\mu'}\rangle\right|^{2}=\frac{\hbar e^{2}\omega_{\mathrm{LO}}}{2\epsilon_{0}V}\left(\frac{1}{\epsilon(\infty)}-\frac{1}{\epsilon(0)}\right)\frac{q_{\lambda}^{2}}{q^{4}}\left|\langle u_{\bvec{k}+\bvec{q}\mu}|u_{\bvec{k}\mu'}\rangle\right|^{2},
	\end{equation}
\end{widetext}
with the elementary charge $e$, the energy of the longitudinal optical
phonons $\omega_{\mathrm{LO}}$, the crystal volume $V$, the vacuum
permittivity $\epsilon_{0}$ and the low and high frequency dielectric
constants $\epsilon(0)$, $\epsilon(\infty)$. The corresponding parameters
can be found in Table \ref{tab:PhonInteraction1}. $q_{\lambda}$ is the corresponding component of the transferred momentum $\bvec{q}$. To obtain this form of the matrix element one has to introduce the simple set of elongation vectors \cite{Scholz1995} 
\begin{align}
	\vec{\varepsilon}^{(1)}_{\lambda\bvec{q}} = \bvec{e}_\lambda \sqrt{\frac{M_1}{M_1 + M_2} }\ ,
	\vec{\varepsilon}^{(2)}_{\lambda\bvec{q}} = -\bvec{e}_\lambda \sqrt{\frac{M_2}{M_1 + M_2} }
\end{align}
with $\lambda \in\{ x,y,z\}$, where $\bvec{e}_\lambda$ is the unit vector along the cubic axes. These elongation vectors are only exact for a nonpolar material, such that in a polar crystal the transversal and longitudinal modes can not be distinguished. Because of this approximation, we have to assume that the optical phonon mode dispersions are nondegenerate with $\hbar \omega_{\mathrm{LO}}\approx\hbar \omega_{\mathrm{TO}} \approx 34.4\ \mathrm{meV}$. This value does not differ a lot from the literature parameters as given in Table \ref{tab:PhonInteraction1}. 

\begin{table}[t]
	\begin{tabular}{|l||c|c|c|c|c|c|c|}
		\hline 
		Quantity & $\hbar\omega_{\mathrm{LO}}$ & $\hbar\omega_{\mathrm{TO}}$ & $c$ & $\epsilon(0)$ & $\epsilon(\infty)$ & $\rho$ & $a_{0}$\tabularnewline
		\hline 
		Value & $36.2$ & $33.3$ & $3860$ & 12.9 & 10.92 & $5316$ & 0.5653\tabularnewline
		\hline 
		Unit & meV & meV & $\mathrm{\frac{m}{s}}$ & 1 & 1 & $\frac{\mathrm{kg}}{\mathrm{m^{3}}}$ & nm\tabularnewline
		\hline 
	\end{tabular}	
	\caption{Material parameters for GaAs hole-phonon interaction, taken from Ref.~\onlinecite{Scholz1995}.\label{tab:PhonInteraction1} }
\end{table}

\begin{table}[t]
	\begin{tabular}{|l||c|c|c|c|}
		\hline 
		Quantity & $d_{0}$ & $a_{v}$ & $b$ & $d$\tabularnewline
		\hline 
		Value & 48 & 1.16 & -1.7 & -4.55\tabularnewline
		\hline 
	\end{tabular}
	\caption{Deformation potential parameters in eV \cite{Chuang1995}. The deformation
		potential $d_{0}$ is taken from \cite{Poetz1981,Grimsditch1979}.
		\label{tab:PhonInteraction2} }
\end{table}

In difference to the Fr\"{o}hlich interaction the short-range deformation
potential interaction is explicit spin dependent and is given by \cite{Scholz1995,Bir1974} 
\begin{equation}
	\hat{v}_{\bvec{k}+\bvec{q}\mu,\bvec{k}\mu'}^{\left(\lambda\right)}=\sqrt{\frac{\hbar}{2\rho V\omega_{\bvec{q},\lambda}}}D_{\bvec{q}}^{\lambda},
	\label{eq:deformation-potential-optical-phonons}
\end{equation}
with the deformation potential matrix 
\begin{equation}
	D_{\bvec{q}}^{\lambda}=\begin{pmatrix}0 & h & j & 0\\
		h^{*} & 0 & 0 & j\\
		j* & 0 & 0 & -h\\
		0 & j^{*} & -h^{*} & 0
	\end{pmatrix},\label{eq:Wechselwirkung mit np op Phononen}
\end{equation}
where the entries are 
\begin{equation}
	\begin{split}h & =\frac{d_{0}}{a_{0}}\left(\delta_{y\lambda}-i\delta_{x\lambda}\right),\\
		j & =\frac{d_{0}}{a_{0}}\delta_{z\lambda}.
	\end{split}
\end{equation}
$\rho$ is the density of GaAs, $d_{0}$ the deformation potential
and $a_{0}$ the lattice constant. The values for the different deformation
potentials, including $d_{0}$ can be found in Table \ref{tab:PhonInteraction2}.
The difference of a factor $i$ compared to \cite{Scholz1995} arises
because of a different definition of the phononic displacement operator.

For the interaction with the acoustic phonons we use
the general form \cite{Bir1974} 
\begin{equation}
	H_{\mathrm{ac}}
	=\begin{pmatrix}F & H & J & 0\\
		H^{*} & G & 0 & J\\
		J* & 0 & G & -H\\
		0 & J^{*} & -H^{*} & F
	\end{pmatrix},
	\label{eq:hamiltonian-deformation-potential-acoustic}
\end{equation}
with 
\begin{equation}
	\begin{split}F & =\frac{l+m}{2}\left(\varepsilon_{xx}+\varepsilon_{yy}\right)+m\varepsilon_{zz},\\
		G & =\frac{1}{3}\{f+2\left[m\left(\varepsilon_{xx}+\varepsilon_{yy}\right)+l\varepsilon_{zz}\right]\},\\
		H & =-\frac{1}{\sqrt{3}}n\left(i\varepsilon_{xz}+\varepsilon_{yz}\right),\\
		J & =\frac{1}{\sqrt{3}}\left[\frac{1}{2}\left(l-m\right)\left(\varepsilon_{xx}-\varepsilon_{yy}\right)-in\varepsilon_{xy}\right],
	\end{split}
\end{equation}
where the constants $l,\ m,\ n$ are defined via the deformation potentials
$a_{v},\ b,\ c$ 
\begin{equation}
	a_{v}=\frac{l+2m}{3},\ b=\frac{l-m}{3},\ d=\frac{n}{\sqrt{3}}.
\end{equation}
The symmetrized strain tensor 
\begin{equation}
	\varepsilon_{\alpha\beta}=\frac{1}{2}\left(\frac{\partial u_{\alpha}}{\partial x_{\beta}}+\frac{\partial u_{\beta}}{\partial x_{\alpha}}\right)
	\label{eq:strain-tensor}
\end{equation}
is evaluated via the phonon displacement operator 
\begin{equation}
	\bvec{u}_{\lambda}(\bvec{x})=i\sqrt{\frac{\hbar}{2\rho V\omega_{\bvec{q},\lambda}}}\sum_{\bvec{q},\lambda}\left(b_{-\bvec{q},\lambda}^{\dagger}+b_{\bvec{q},\lambda}\right)e^{i\bvec{q}\cdot\bvec{x}}\vec{\varepsilon}_{\lambda}(\bvec{q})
\end{equation}
where $\lambda$ runs over the modes of the acoustic phonons. We use the elongation vectors given in Pikus and Bir~\cite{Bir1974}: $\vec{\varepsilon}_{\text{LA}}(\bvec{q})=\bvec{q}/{q},$ 
\begin{equation}
	\vec{\varepsilon}_{\text{TA}1}(\bvec{q})=\frac{1}{q_{\perp}}
	\begin{pmatrix}
		q_{y}\\
		q_{x}\\
		0
	\end{pmatrix},\ 
	\vec{\varepsilon}_{\text{TA}2}(\bvec{q})=\frac{1}{q q_{\perp}}
	\begin{pmatrix}
		q_{x}q_{z}\\
		q_{y}q_{z}\\
		-q_{\perp}^{2}
	\end{pmatrix}.
\end{equation}
For the dispersion, we assume $\hbar\omega_{\bvec{q},\lambda}=\hbar c q$,
with sound velocity $c$. We find for the deformation
potential interaction 
\begin{equation}
	\hat{v}_{\bvec{k}+\bvec{q}\mu,\bvec{k}\mu'}^{\left(\lambda\right)}=\sqrt{\frac{\hbar}{2\rho V\omega_{\bvec{q},\lambda}}}D_{\bvec{q}}^{\lambda},
	\label{eq:acoustic-matrixelement-D}
\end{equation}
with
\begin{equation}
	D_{\bvec{q}}^\mathrm{\lambda} = 
	\begin{pmatrix}
		f & h & j & 0\\  
		h^* & g & 0 & j\\
		j* & 0 & g & -h\\
		0 & j^* & -h^* & f
	\end{pmatrix},
\end{equation}
where the entries are for the longitudinal mode
\begin{align}
	\begin{split}
		f&=\frac{l+m}{2q}\left( q_x^2+q_y^2\right)+\frac{m}{q}q_z^2,\\ g&=\frac{1}{3}\{f+\frac{2}{q}\left[m\left(q_x^2+q_y^2\right)+lq_z^2]\right]\}, \\
		h&=-\frac{q_z}{q\sqrt{3}}n\left(iq_x+q_y\right),\\
		j&=\frac{1}{q\sqrt{3}}\left[ \frac{1}{2} \left( l-m\right) \left(q_x^2-q_y^2 \right)-i n q_x q_y\right]
	\end{split}
\end{align}
and for the two transversal modes
\begin{align}
	\begin{split}
		f&=0,\\ 
		g&=0, \\
		h&=-\frac{q_z}{2q_\perp\sqrt{3}}n\left(iq_y-q_x\right),\\ 
		j&=\frac{1}{q_\perp\sqrt{3}}\left[   \left( l-m\right) q_y q_x-i \frac{n}{2} \left( q_y^2-q_x^2\right)\right],
	\end{split}
\end{align}
and
\begin{align}
	\begin{split}
		f&=\frac{l-m}{2q q_\perp}q_z\left( q_x^2+q_y^2\right),\\
		g&=-f,\\
		h&=-\frac{1}{2qq_\perp\sqrt{3}}n\left[i q_x\left(q_z^2-q_x^2-q_y^2\right)+q_y\left(q_z^2-q_x^2-q_y^2\right)\right], \\
		j&=\frac{1}{q q_\perp\sqrt{3}}\left[ \frac{1}{2} \left( l-m\right) q_z \left(q_x^2-q_y^2 \right)-i n q_x q_y q_z\right].
	\end{split}
\end{align}

\end{appendix}

\bibliographystyle{apsrev4-1}
%

\end{document}